\documentclass[aps,prb,amsmath,amssymb,reprint,longbibliography]{revtex4-1}
\usepackage{amssymb}
\usepackage{amsmath}
\usepackage{graphicx}
\usepackage{color}
\usepackage{chemformula}
\usepackage{natbib}
\usepackage{subcaption}
\usepackage{upgreek}
\usepackage{verbatim}

\begin{document}
\title{Machine-learning Based Screening of Lead-free Halide Double Perovskites for Photovoltaic Applications}
\author{Elisabetta Landini}
\affiliation{Fritz-Haber-Institut der Max-Planck-Gesellschaft, Faradayweg 4-6, D-14195 Berlin, Germany}
\affiliation{Chair for Theoretical Chemistry and Catalysis Research Center, Department of Chemistry, Technische Universit\"at M\"unchen, Lichtenbergstr. 4, 85748 Garching, Germany}
\author{Karsten Reuter}
\affiliation{Fritz-Haber-Institut der Max-Planck-Gesellschaft, Faradayweg 4-6, D-14195 Berlin, Germany}
\author{Harald Oberhofer}
\affiliation{Chair for Theoretical Physics VII and Bavarian Center for Battery Technology (BayBatt), University of Bayreuth, 95447 Bayreuth, Germany}
\email{harald.oberhofer@uni-bayreuth.de}

\begin{abstract}
Lead-free halide double perovskites are promising stable and non-toxic alternatives to methylammonium lead iodide in the field of photovoltaics. In this context, the most commonly used double perovskite is Cs$_2$AgBiBr$_6$, due to its favorable charge transport properties.
However, the maximum power conversion efficiency obtained for this material does not exceed 3\%, as a consequence of its wide indirect gap and its intrinsic and extrinsic defects.  On the other hand, the materials space that arises from the substitution of different elements in the 4 lattice sites of this structure is large and still mostly unexplored.
In this work a neural network is used to predict the band gap of double perovskites  from an initial space of 7056 structures and select candidates suitable for visible light absorption. Successive hybrid DFT calculations are used to evaluate the thermodynamic stability, the power conversion efficiency and the effective masses of the selected compounds, and to propose novel potential solar absorbers.
\end{abstract}
\maketitle

\section{Introduction}
In the last years, perovskite solar cells have attracted significant attention in the field of photovoltaics, due to their low production cost and high efficiency. The absorber material in these devices is a hybrid organic-inorganic perovskite containing an organic cation (e.g.~methylammonium or formamidinium), lead and a halide\cite{hocker2021temperature,biewald2019temperature,leguy2015reversible,tress2017perovskite,zhao2016organic}
. Methylammonium lead iodide (MAPI) based solar cells  were first reported in 2009 with an efficiency of 3.8\%\cite{kojima2009organometal} and their performance rapidly increased in the following years. Today, the highest power conversion efficiency for a perovskite solar cell is 25.5\%\cite{green2021solar} for formamidinium lead iodide (FAPI) thin films, which is comparable to that of Si single crystal solar cells. These materials posses ideal charge transport properties such as high and balanced carrier mobility and long diffusion length \cite{bi2016charge,dittrich2016diffusion}, but the toxicity of Pb and the instability of the structure when exposed to humidity and heat prevent a large scale application of this technology. Several attempts have been made to replace Pb\textsuperscript{2+} with Ge\textsuperscript{2+} or Sn\textsuperscript{2+}, however, the resulting structures have shown poor stability due to the oxidation of Ge and Sn to 4+ formal charge state\cite{ogomi2014ch3nh3sn,noel2014lead}. 
Another possible strategy is to substitute two Pb atoms with two cations of charge +1 and +3 in an alternated way, to form the double perovskite structure \ch{A2BB'X6}. Some of the most extensively studied halide double perovskites in the field of photovoltaics are \ch{Cs2AgBiBr6}\cite{volonakis2016lead,filip2016band,slavney2017defect,mcclure2016cs2agbix6,bartesaghi2018charge,sirtl2020optoelectronic,yang2020simultaneous,armer2021influence} and \ch{Cs2AgInCl6}\cite{luo2018cs2agincl6,dahl2019probing}. However, solar cells based on the former have never reached an efficiency above 3\% due to its large indirect gap, while the latter has a direct gap but the parity-forbidden transition\cite{meng2017parity} makes it a weak absorber. A small direct gap of 0.95 eV was found for \ch{Cs2AgTlBr6}\cite{slavney2018small} but the high toxicity of Tl limits its application.

Given that double perovskites are quaternary compounds, there are in principle many thousands of possible compositions, with an equally large potential range of optoelectronic properties. Sampling this large design space, though, is not necessarily an easy task even for theory. Considering that accurate predictions of the perovskite electronic structure necessitate computationally expensive theoretical methods such as hybrid functionals and spin-orbit coupling\cite{filip2016band}. Earlier studies of perovskites, performing e.g.~high-thoughput screening,\cite{jacobs2019materials,tao2017accurate} were thus often limited in the employed methods, the search range, or focused only on a subset of properties.

Notable examples here include the works of \citeauthor{filip2018phase}\cite{filip2018phase}, \citeauthor{zhao2017design}\cite{zhao2017design}, \citeauthor{volonakis2017route}\cite{volonakis2017route}, \citeauthor{roknuzzaman2019electronic}\cite{roknuzzaman2019electronic}, and \citeauthor{ding2022high}\cite{ding2022high}, which screened limited search spaces up to a thousand compounds. These studies employed density functional theory at the semi-local, or sometimes at the hybrid level, to also compute the compounds stabilities. Larger studies of up to a few thousand double perovskites,\cite{cai2019high,bartel2020inorganic} on the other hand, tended to employ geometric arguments\cite{bartel2019new} to pre-filter the search space.

Recognizing the limits of first-principles calculations, several groups employed machine learning (ML)-based methods to choose potential candidate materials. Unfortunately, such methods tend to also need large amounts of data for their training, which often was generated using comparatively cheap computational methods.

\citeauthor{schmidt2017predicting}\cite{schmidt2017predicting} generated a database of almost 250,000 cubic perovskites calculated via DFT with the PBE(+U) functional. Among these, 641 were found to be stable and overall 1562 had a PBE gap above 0.5 eV. This database was also used as benchmark to compare the performance of several ML models (ridge regression, neural networks, random forest, and extremely randomized trees) on the prediction of the energy above hull . The most accurate method was found to be extremely randomized trees. \citeauthor{saidi2020machine}\cite{saidi2020machine} used a hierarchical convolutional neural network (CNN) to predict the lattice constant, octahedral angle and band gap of hybrid metal halide perovskites \ch{ABX3}, focusing mainly on the effect of the organic cation A. \citeauthor{li2020progressive}\cite{li2020progressive} trained several ML models (gradient tree boosting regression, kernel ridge regression, support vector regression, bootstrap aggregating regression, Gaussian process regression (GPR) and random forest) to predict the formation energy of \ch{ABO3} oxide perovskites and used it as instrumental variable to successively predict the band gap. The models were trained on 1593 oxide perovskites obtained from the Materials Project database\cite{jain2013commentary} and resulted in a lowest mean absolute error of 0.384 eV for the GPR model.
\citeauthor{pilania2016machine}\cite{pilania2016machine} applied kernel ridge regression (KRR) to predict the band gap  of oxide double perovskites (AA'BB'O$_6$), starting from a database of 1306 DFT band gaps calculated with the GLLB-SC functional\cite{kuisma2010kohn}. In their workflow, they selected 16 elemental features and 16 LASSO-based\cite{nelson2013compressive} compound features to build descriptors to be tested via a linear least square fit (LLSF). Subsequently, they applied KRR on the best-performing ones, obtaining a final root mean squared error (RMSE) of 0.36 eV on a test set including 10\% of the structures in the database.
\citeauthor{l2019machine}\cite{l2019machine} trained a random forest algorithm on the band gaps of chalcogenide double perovskites calculated via hybrid DFT using the HSE06 functional to identify promising solar absorbers. The stability of the selected compounds was evaluated by combining information from the geometric Goldschmidt tolerance factor, the decomposition energy and molecular dynamics (MD) simulations. For the stable compounds, optical absorption was calculated, leading to the discovery of 5 promising sulfide double perovskites.
In the field of halide double perovskites, \citeauthor{konno2020deep}\cite{konno2020deep} employed a convolutional neural network to implicitly extract elemental features from the position of the atoms in the periodic table. The CNN was trained and tested on a dataset of 3734 experimental band gaps and predicted band gaps with a RMSE of 0.42 eV. \citeauthor{im2019identifying}\cite{im2019identifying} used a gradient boosting regression tree (GBRT) to predict the formation energy and the band gap of halide double perovskites, reaching a RMSE of 0.021 eV/atom and 0.223 eV respectively. The model was trained on the DFT data of 540 compounds calculated with the PBE functional. \citeauthor{yang2021machine}\cite{yang2021machine} compared the performance of GBRT, ridge regression, support
vector regression, KRR, a bagging ensemble algorithm, and a random forest ensemble algorithm and applied GBRT to explore an initial space of 16400 double perovskites.

\begin{figure}[htb]
\centering
    \includegraphics[width=\columnwidth]{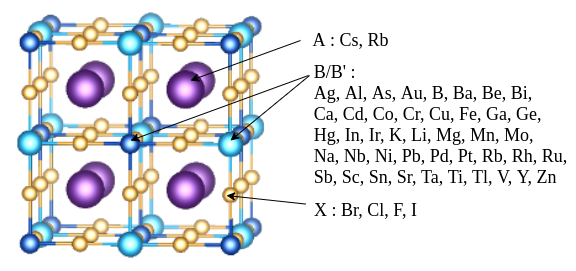}
\caption{Double perovskite structure and the compositional space explored}  
\label{fig:perovsk}
\end{figure}

Following in these footsteps, we here use a convolutional neural network with a periodic table representation (PTR) of the input compounds,\cite{zheng2018machine} to sample a space of 7056 double perovskites with 2 alkali metals in the position A, 44 metals in the position B/B' and 4 halides in the position X (Figure~\ref{fig:perovsk}). The PTR has shown to be very useful in any material's discovery situation where the elements of the search space share a similar basic structure, such as Heusler compounds\cite{zheng2018machine} or, as in our case, double perovskites\cite{konno2021deep}. It also has the great advantage that no additional computations are necessary to generate\cite{konno2020deep}. Finally the PTR essentially represents the material as a periodic 2D image and thus lends itself naturally to a treatment by CNNs which were, after all, conceived for image recognition purposes.

In this work we focus on a comprehensive set of properties relevant for photovoltaic applications, the thermodynamic stability of the materials, their power conversion efficiency and the effective masses of their charge carriers.\cite{muschielok2019aspects} All our training and test data are computed on the level of hybrid DFT including spin-orbit coupling to ensure a high predictivity of our results. We find a number of promising perovskites not yet considered in experiment, with power conversion efficiencies above 15\%. Upon relaxing the criterion of cubic symmetry, we also find a number of perovskite-like structures, with equally high efficiencies but better predicted thermodynamic stabilities.

\section{Methods}
The training and test sets include 764 and 200 structures, respectively, calculated by DFT using the FHI-aims code\cite{blum2009ab}. We used the HSE06\cite{heyd2003hybrid,krukau2006influence} exchange-correlation functional, a (4x4x4) k-point grid, an energy convergence threshold of 10$^{-6}$ eV and a density convergence threshold of 10$^{-6}$ e/a$_0^3$, where a$_0$ is a Bohr radius. We use the numeric atom-centered \textit{light} basis sets as implemented in FHI-aims\cite{blum2009ab}. For most computations, the geometry was optimized in a symmetry-preserving framework, with a covergence threshold for the forces of 10$^{-2}$ eV/\AA. The calculations include a non-self-consistent spin-orbit coupling correction\cite{huhn2017one} and collinear treatment of the spin, which was initialized following the configuration of isolated atoms. In the structures containing two metals with unpaired electrons, the two spins were initialized in a antiparallel configuration. 

The input for the convolutional neural network is a tensor with 3 channels, one for each of the A, B, and X sites. Each channel has the shape of a periodic table filled with zeros and with the stoichiometric coefficient (2 for A, 1 for B and B' and 6 for X) on the position of the four elements in the structure.
The neural network (cf.~Table~\ref{tab:network}) has 5 convolutional layers with four 2x2 kernels and one 1x2 kernel, and 4 fully connected layers. Each layer has LeakyReLU\cite{maas2013rectifier} as activation function, with a negative slope of 0.2. The fully connected layers include layer normalization\cite{ba2016layer} and dropout\cite{hinton2012improving} with a probability of 0.25. The network was trained using the Adam optimizer\cite{kingma2014adam} on batches of 100 samples for 1000 training set iterations, with a learning rate of 0.001, a weight decay of 0.0005 and cosine annealing\cite{loshchilov2016sgdr} as learning rate scheduler. The hyperparameters were chosen in order to minimize the loss between predicted and calculated band gap values on the test set.
For the structures with a predicted band gap between 0.9 and 1.6 eV, the the thermodynamic stability was estimated by calculating the enthalpy of the decomposition reaction.\cite{Pu2021screening}
For the stable compounds the absorption coefficient was calculated using the random phase approximation on an increased k-point grid density of (8x8x8) and a gaussian broadening of 0.05 eV. From this, the spectroscopic limited maximum efficiency (SLME) was calculated to estimate the maximum theoretical power conversion efficiency, following a method proposed by \citeauthor{yu2012identification}\cite{yu2012identification}. Additionally, the effective mass was calculated by parabolic fit of the band edges along the high symmetry directions.
\begin{table}[ht]
    \centering
    \begin{tabular}{ccccc}
        \hline
        \multicolumn{5}{c}{Convolutional layers} \\
        \hline
        Kernel & Channels & Activation & Dropout & Norm\\
        \hline
        2x2 & 3,100 & LeakyReLU(0.2) & 0.25 & -\\
        2x2 & 100,100 & LeakyReLU(0.2) & 0.25 & - \\
        2x2 & 100,100 & LeakyReLU(0.2) & 0.25 & - \\
        2x2 & 100,100 & LeakyReLU(0.2) & 0.25 & - \\
        1x2 & 100,100 & LeakyReLU(0.2) & 0.25 & - \\
        \hline
        \multicolumn{5}{c}{Fully connected layers} \\
        \hline
        \multicolumn{2}{c}{Nodes} & Activation & Dropout & Norm\\
        \hline
        \multicolumn{2}{c}{1200,200} & LeakyReLU(0.2) & 0.25 & layer\\
        \multicolumn{2}{c}{200,200} & LeakyReLU(0.2) & 0.25 & layer\\
        \multicolumn{2}{c}{200,200} & LeakyReLU(0.2) & 0.25 & layer\\
        \multicolumn{2}{c}{200,1} & LeakyReLU(0.2) & 0.25 & layer\\
        \hline
    \end{tabular}
    \caption{Network architecture}
    \label{tab:network}
\end{table}

\section{Results and discussion}
\begin{figure}[htb]
\centering
    \includegraphics[width=\columnwidth]{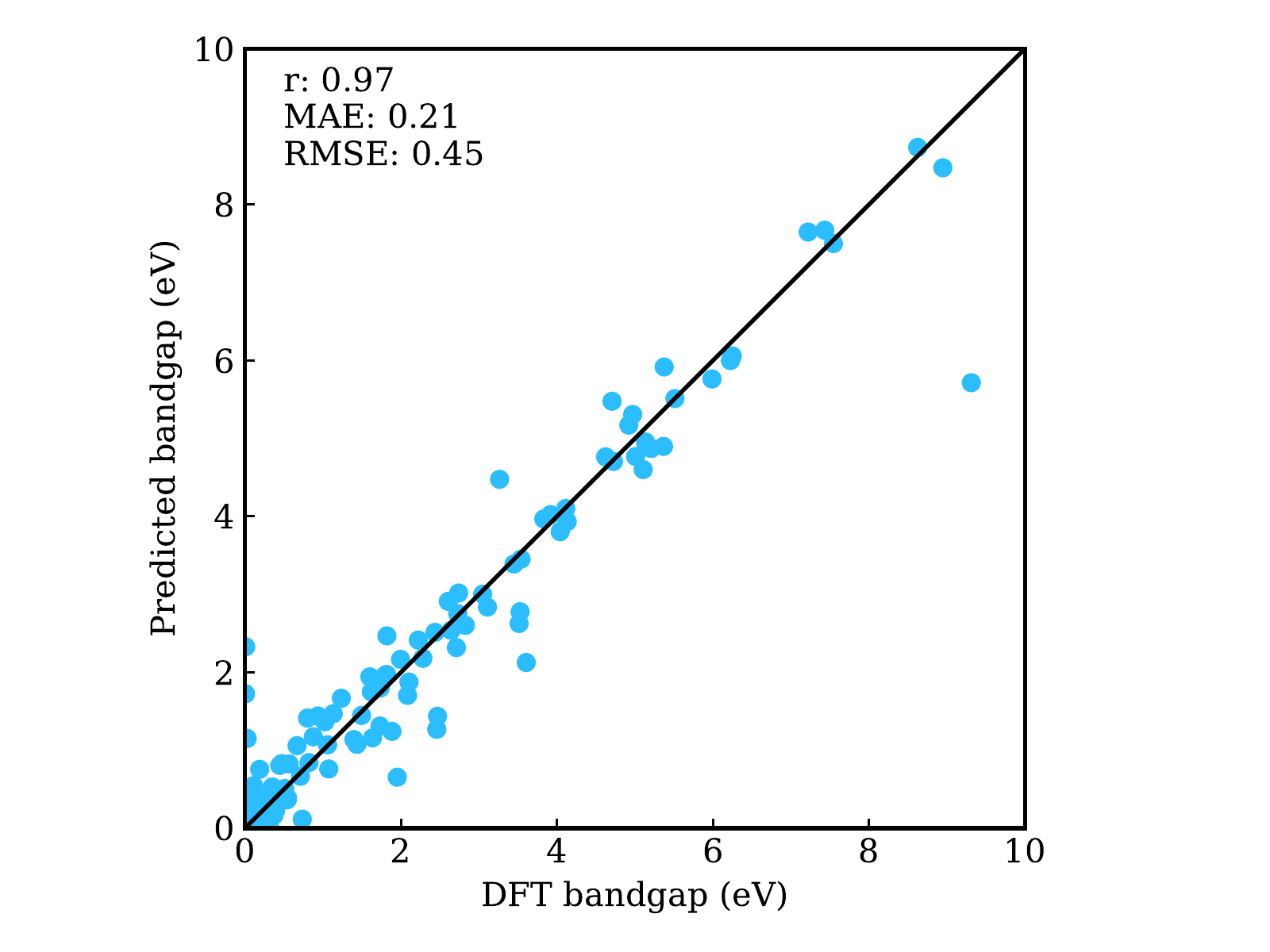}
\caption{Training result on a test set of 200 structures}
\label{fig:trained_model}
\end{figure}

\begin{figure}[htb]
\centering
    \includegraphics[width=\columnwidth]{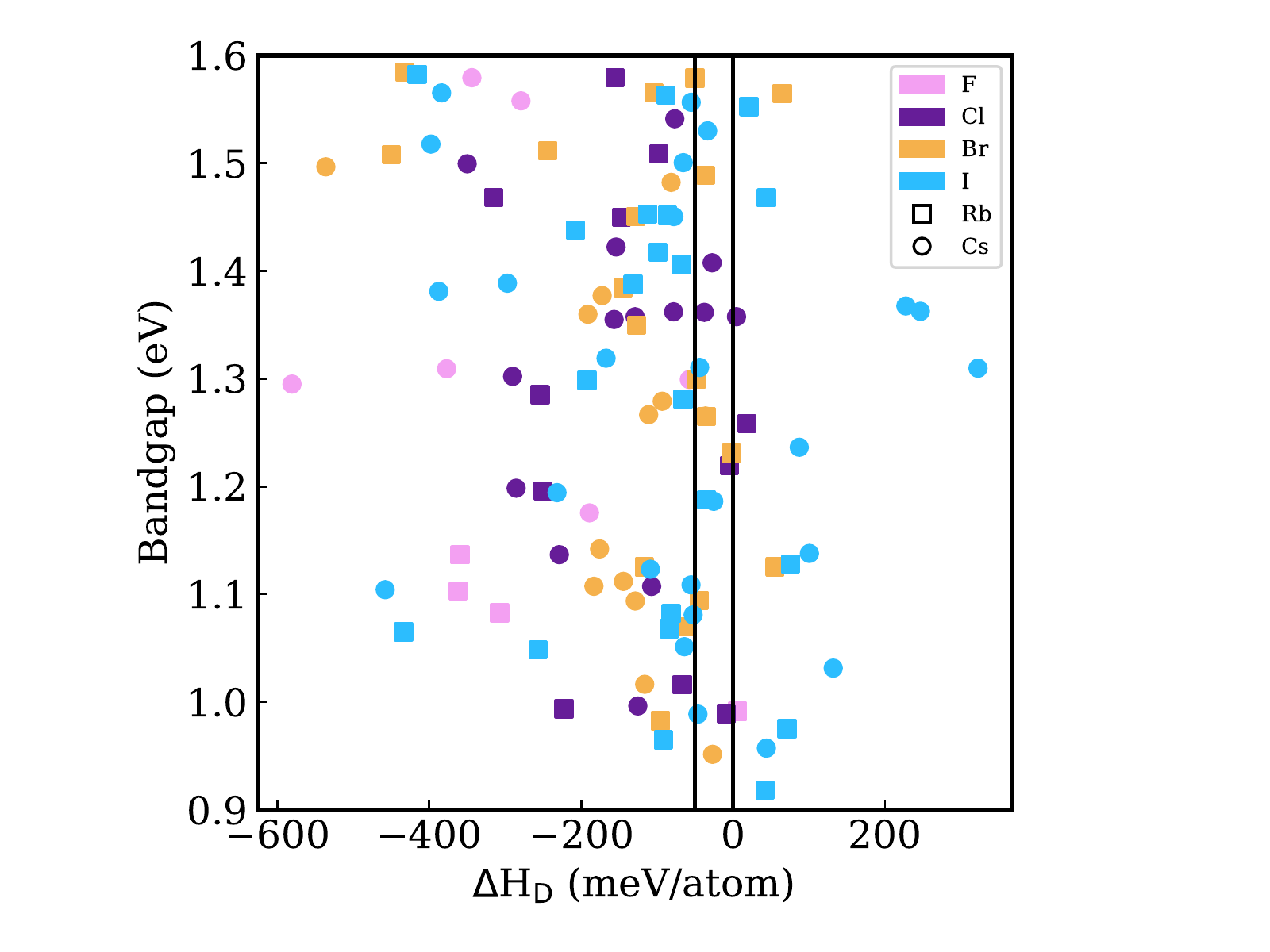}
\caption{Enthaply of the decomposition reaction as a function of DFT band gap. The compounds with $\Delta$H\textsubscript{D}$>$0 meV/atom are stable, while those included between the vertical lines (from -50 to 0 meV/atom) are assumed to be metastable (see text)}  
\label{fig:decomposition_enthalpy}
\end{figure}

The result of the training on a randomly sampled set of 200 structures is shown in Figure~\ref{fig:trained_model}. With the exception of some outliers the model can predict the band gap with a reasonable accuracy and an overall MAE of 0.21 eV and RMSE of 0.45 eV. The machine learning model is then used to select candidates with a predicted band gap between 0.9 and 1.6 eV. From this set of 459 structures, first those that contain toxic elements have been excluded. Subsequently, the band gap of the remaining 303 structures was explicitly computed at the hybrid DFT level (unless they were already contained in the initial training or test set for the ML model). For the 119 compounds with a DFT band gap included in the same interval, the decomposition enthalpy has been calculated (cf.~Figure~\ref{fig:decomposition_enthalpy}). For a compound to be stable the decomposition enthalpy must be positive, but given the finite accuracy of the employed approximate DFT functional and the fact that the DFT calculations don't include any temperature effect, also the structures with a decomposition enthalpy between -50 and 0 meV/atom have been included in the successive analysis -- as candidates potentially stable at room temperature or metastable. As it has been shown by \citeauthor{sun2016thermodynamic}\cite{sun2016thermodynamic}, metastable compounds are often found in this energy interval.


\begin{figure}[htb]
\centering
    \includegraphics[width=\columnwidth]{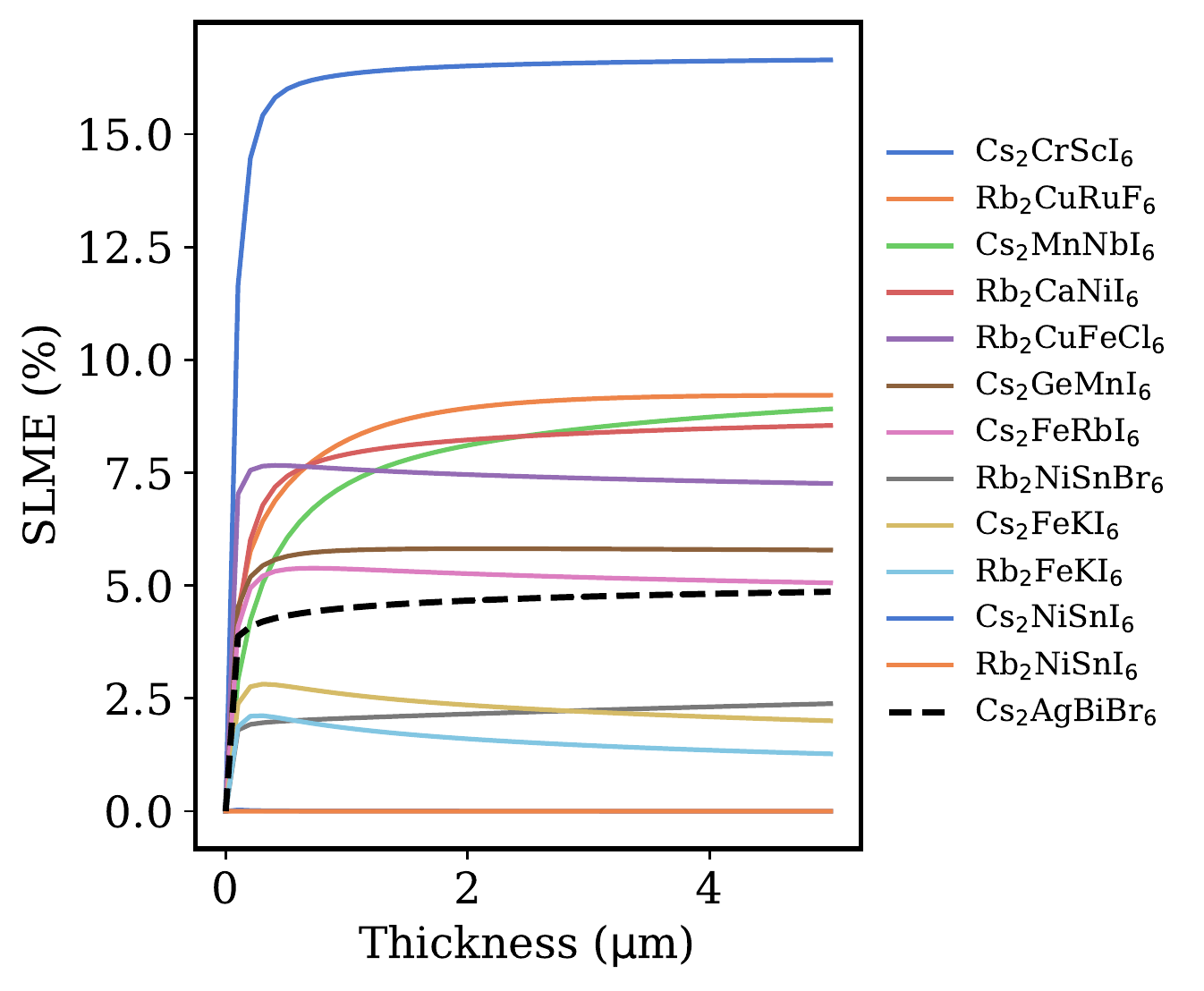}
\caption{SLME for stable structures. The efficiency of \ch{Cs2AgBiBr6} is shown for comparison}
\label{fig:slme_stable}
\end{figure}

\begin{figure}[htb]
\centering
    \includegraphics[width=\columnwidth]{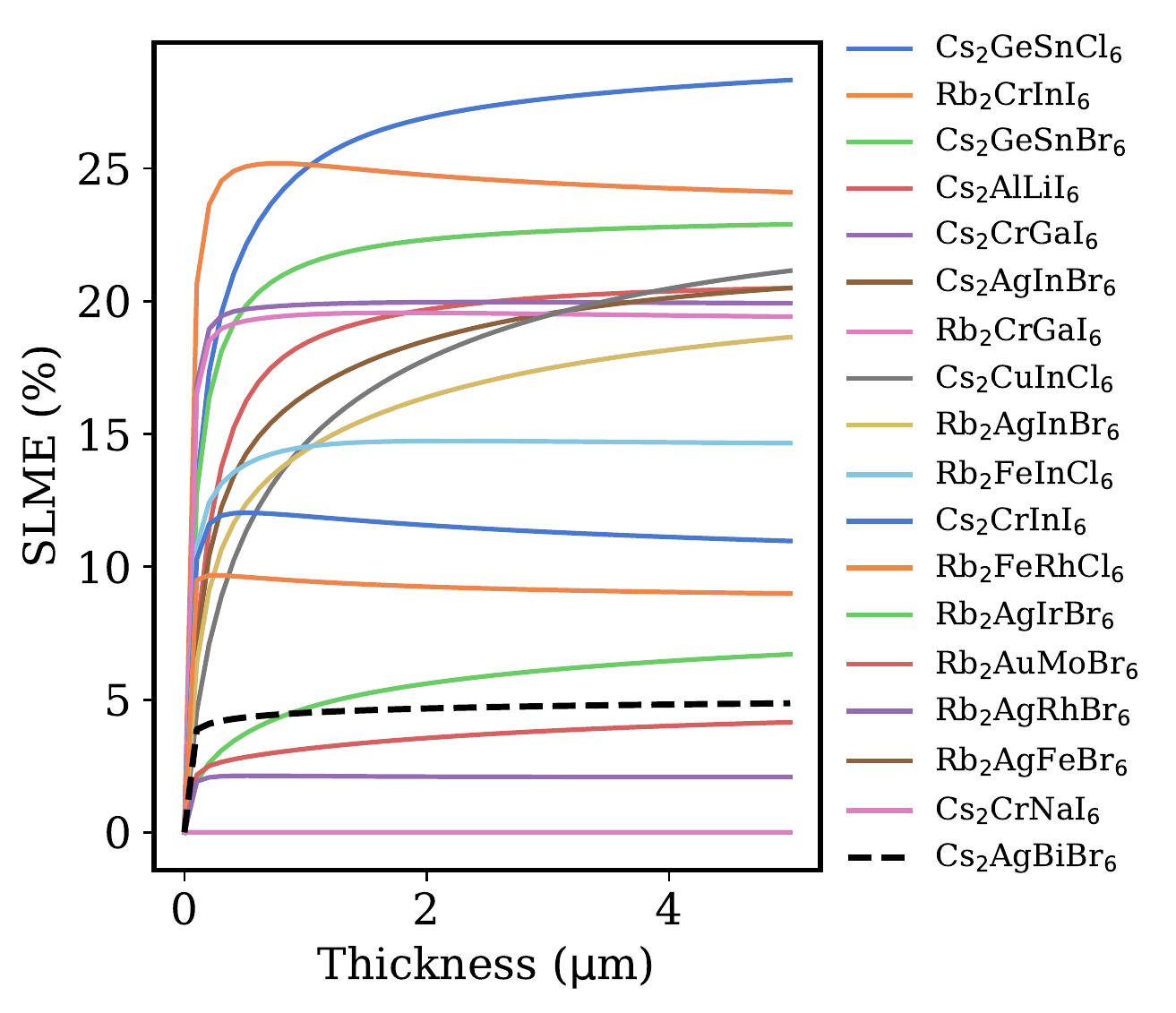}
\caption{SLME for metastable structures. The efficiency of \ch{Cs2AgBiBr6} is shown for comparison}
\label{fig:slme_metastable}
\end{figure}

The maximum theoretical power conversion efficiency of the 12 stable and 17 metastable candidates thus identified was calculated for a layer thickness up to 5 $\upmu$m (see Figure~\ref{fig:slme_stable}~and~\ref{fig:slme_metastable}).


\begin{table*}[htb]
    \centering
    \begin{tabular}{r|cccccc}
\hline
Structure & $\Delta$H\textsubscript{D} (meV/at) & E\textsubscript{gap} (eV) & E\textsubscript{abs} (eV) & SLME (\%) &
m$^*_h$ (m$_0$) & m$^*_e$ (m$_0$)\\
\hline
\ch{Rb2AgIrBr6} & -48 &  1.30 &  2.03 &  6 & -1.71 (X-$\Gamma$) -3.81 (X-W) & 0.34 (X-$\Gamma$) 0.31 (X-W)\\
 \ch{Cs2CrNaI6} & -46 &  0.99 &  2.20 &  0 & -1.88 ($\Gamma$-X) -0.95 ($\Gamma$-L) & 0.49 (X-$\Gamma$) 0.52 (X-W)\\
\ch{Rb2AgFeBr6} & -45 &  1.09 &  2.31 &  0 & -0.38 ($\Gamma$-X) -0.52 ($\Gamma$-L) & 0.56 (X-$\Gamma$) 0.46 (X-W)\\
 \ch{Cs2AlLiI6} & -44 &  1.31 &  1.67 & 20 & -0.81 ($\Gamma$-X) -1.24 ($\Gamma$-L) & 0.28 ($\Gamma$-X) 0.28 ($\Gamma$-L)\\
 \ch{Rb2CrInI6} & -43 &  0.97 &  1.00 & 24 & -0.42 (L-$\Gamma$) -0.14 (L-W) & 0.65 (L-$\Gamma$) 0.19 (L-W)\\ 
\ch{Cs2CuInCl6} & -38 &  1.36 &  1.65 & 19 & * ($\Gamma$-X) -1.06 ($\Gamma$-L) & 0.27 ($\Gamma$-X) 0.27 ($\Gamma$-L)\\
\ch{Cs2AgInBr6} & -36 &  1.27 &  1.61 & 20 & -5.62 ($\Gamma$-X) -0.71 ($\Gamma$-L) & 0.20 ($\Gamma$-X) 0.20 ($\Gamma$-L)\\
\ch{Rb2AgRhBr6} & -36 &  1.49 &  2.67 &  2 & -1.06 (L-$\Gamma$) -0.69 (L-W) & 0.36 (X-$\Gamma$) 0.30 (X-W) \\
 \ch{Rb2CrGaI6} & -35 &  1.19 &  1.49 & 20 & -1.26 ($\Gamma$-X) -0.86 ($\Gamma$-L) & 0.66 (L-$\Gamma$) 0.26 (L-W)\\
\ch{Rb2AuMoBr6} & -35 &  1.26 &  2.15 &  4 & * & *\\
 \ch{Cs2CrInI6} & -29 &  1.03 &  1.38 & 11 & -0.45 (L-$\Gamma$) -0.15 (L-$\Gamma$) & 0.66 (L-W) 0.21 (L-W)\\
\ch{Cs2GeSnCl6} & -28 &  1.41 &  1.41 & 28 & -0.13 ($\Gamma$-X) -0.13 ($\Gamma$-L) & 0.20 ($\Gamma$-X) 0.20 ($\Gamma$-L)\\ 
\ch{Cs2GeSnBr6} & -27 &  0.95 &  0.95 & 23 & -0.09 ($\Gamma$-X) -0.09 ($\Gamma$-L) & 0.14 ($\Gamma$-X) 0.14 ($\Gamma$-L)\\ 
 \ch{Cs2CrGaI6} & -26 &  1.19 &  1.49 & 20 & -1.22 ($\Gamma$-X) -0.88 ($\Gamma$-L) & 0.67 (L-$\Gamma$) 0.27 (L-W)\\
\ch{Rb2FeRhCl6} &  -9 &  0.99 &  1.58 &  9 & -0.39 (L-$\Gamma$) -0.36 (L-W) & 1.80 (X-$\Gamma$) 0.59 (X-W)\\
\ch{Rb2FeInCl6} &  -4 &  1.19 &  1.73 & 15 & -0.29 (X-$\Gamma$) -1.35 (X-W) & 27.30 (L-$\Gamma$) 8.31 (L-W)\\
\ch{Rb2AgInBr6} &  -2 &  1.23 &  1.58 & 17 & -7.39 ($\Gamma$-X) -0.66 ($\Gamma$-L) & 0.20 ($\Gamma$-X) 0.20 ($\Gamma$-L)\\
\hline
 \ch{Rb2CuRuF6} &   6 &  0.99 &  1.44 &  9 & -1.15 ($\Gamma$-X) -1.59 ($\Gamma$-L) & 15.00 (L-$\Gamma$) 7.60 (L-W)\\ 
 \ch{Rb2CuFeCl6} &  18 &  1.26 &  2.05 &  7 &  * ($\Gamma$-X) -0.79 ($\Gamma$-L) & 0.78 (X-$\Gamma$) 0.62 (X-W)\\ 
 \ch{Rb2NiSnI6} &  42 &  0.92 &  1.96 &  0 & -0.20 (X-$\Gamma$) -0.99 (X-W) & 0.80 (L-$\Gamma$) 0.40 (L-W)\\
 \ch{Rb2CaNiI6} &  44 &  1.47 &  2.33 &  8 & -1.20 ($\Gamma$-X) -0.85 ($\Gamma$-L) & 0.69 (X-$\Gamma$) 0.92 (X-W)\\
 \ch{Cs2NiSnI6} &  44 &  0.96 &  1.93 &  0 & -0.21 (X-$\Gamma$) -1.09 (X-W) & 0.82 (L-$\Gamma$) 0.40 (L-W) \\
\ch{Rb2NiSnBr6} &  65 &  1.56 &  2.72 &  2 & -0.22 (X-$\Gamma$) -1.18 (X-W) & 0.99 (L-$\Gamma$) 0.50 (L-W) \\
  \ch{Rb2FeKI6} &  76 &  1.13 &  2.13 &  1 & -0.17 ($\Gamma$-X) -1.14 ($\Gamma$-L) & 0.55 (X-$\Gamma$) 0.80 (X-W)\\
 \ch{Cs2CrScI6} &  77 &  1.37 &  1.88 & 17 & -0.87 (L-$\Gamma$) -0.55 (L-W) & 0.54 (X-$\Gamma$) 0.46 (X-W)\\ 
 \ch{Cs2FeRbI6} &  87 &  1.24 &  2.13 &  5 & -1.97 ($\Gamma$-X) -1.46 ($\Gamma$-L) & 8.25 (X-$\Gamma$) 1.17 (X-W) \\
  \ch{Cs2FeKI6} & 100 &  1.14 &  2.10 &  2 & -1.56 ($\Gamma$-X) -1.21 ($\Gamma$-L) & 5.72 (X-$\Gamma$) 1.63 (X-W) \\
 \ch{Cs2MnNbI6} & 246 &  1.36 &  2.16 &  8 & * & * \\ %
 \ch{Cs2GeMnI6} & 322 &  1.31 &  2.43 &  6 & -0.22 (X-$\Gamma$) -0.87 (X-W) & 0.27 (L-$\Gamma$) 0.20 (L-W) \\
\hline
\end{tabular}
    \caption{Decomposition enthalpy, DFT band gap, absorption energy, spectroscopic limited maximum power conversion efficiency at 5$\upmu$m and carrier effective masses of the identified metastable and stable compounds, after symmetry constrained geometry optimization. Missing values (*) are due to one or both band edges having bands too narrow to calculate the curvature.}
    \label{tab:cubic}
\end{table*}

\begin{table}[ht]
    \centering
    \begin{tabular}{r|ccccc}
\hline
Structure & $\Delta$H\textsubscript{D} & E\textsubscript{gap} & E\textsubscript{abs} & SLME & Spacegroup\\
          & (meV/atom) & (eV) & (eV) & (\%) & symbol\\
\hline
 \ch{Rb2AgIrBr6} & -32 &  1.95 &  2.43 & 15 & I4\\
 \ch{Cs2CrNaI6} & -44 &  0.99 &  1.63 &  8 & Fm-3m\\
\ch{Rb2AgFeBr6} & -43 &  1.04 &  2.31 &  0 & Fm-3m\\
 \ch{Cs2AlLiI6} & -42 &  1.30 &  1.67 & 20 & Fm-3m\\
 \ch{Rb2CrInI6} & -42 &  0.91 &  0.93 & 23 & Fm-3m\\
 \ch{Cs2CuInCl6} & -28 &  2.21 &  2.54 & 12 & I4/mmm\\
\ch{Cs2AgInBr6} & -34 &  1.22 &  1.58 & 18 & Fm-3m\\
\ch{Rb2AgRhBr6} & -33 &  1.45 &  2.27 &  8 & Fm-3m\\
 \ch{Rb2CrGaI6} & -34 &  1.19 &  1.47 & 19 & Fm-3m\\
 \ch{Rb2AuMoBr6} &   6 &  2.69 &  3.27 &  4 & P-1\\
 \ch{Cs2CrInI6} & -28 &  0.97 &  1.00 & 24 & Fm-3m\\
\ch{Cs2GeSnCl6} & -27 &  1.35 &  1.35 & 27 & Fm-3m\\
\ch{Cs2GeSnBr6} & -26 &  0.89 &  0.90 & 21 & Fm-3m\\
 \ch{Cs2CrGaI6} & -24 &  1.21 &  1.48 & 21 & Fm-3m\\
\ch{Rb2FeRhCl6} &  -8 &  0.93 &  1.54 &  10 & Fm-3m\\
\ch{Rb2AgInBr6} &  -0.7 &  1.20 &  1.57 & 17 & Fm-3m\\
\ch{Rb2FeInCl6} &  -4 &  1.06 &  1.61 & 14 & Fm-3m\\
\hline
 \ch{Rb2CuRuF6} &   7 &  0.91 &  1.40 &  6 & I4/mmm\\ 
\ch{Rb2CuFeCl6} &  19 &  1.13 &  2.10 &  4 & I4/mmm\\
 \ch{Rb2NiSnI6} &  43 &  0.87 &  1.97 &  0 & Fm-3m\\
 \ch{Rb2CaNiI6} &  44 &  1.48 &  2.35 &  8 & Fm-3m\\
 \ch{Cs2NiSnI6} &  45 &  0.92 &  1.94 &  0 & Fm-3m\\
\ch{Rb2NiSnBr6} &  66 &  1.51 &  2.72 &  1 & Fm-3m\\
  \ch{Rb2FeKI6} &  77 &  1.13 &  2.15 &  1 & Fm-3m\\
 \ch{Cs2CrScI6} &  79 &  1.37 &  1.88 & 17 & Fm-3m\\
  \ch{Cs2FeRbI6} & 105 &  1.40 &  2.13 & 11 & C2/m\\
  \ch{Cs2FeKI6} & 102 &  1.14 &  2.13 &  2 & Fm-3m\\
 \ch{Cs2MnNbI6} & 248 &  1.41 &  2.17 &  9 & Fm-3m\\
 \ch{Cs2GeMnI6} & 323 &  1.29 &  2.43 &  6 & Fm-3m\\
\hline
\end{tabular}
    \caption{Decomposition enthalpy, DFT band gap, absorption energy, spectroscopic limited maximum power conversion efficiency at 5$\upmu$m of the metastable and stable compounds, after non symmetry constrained geometry optimization}
    \label{tab:relaxed}
\end{table}

The results are summarized in Table~\ref{tab:cubic}. 

Efficiencies above 20\% were found for compounds with a dipole-allowed transition in correspondence of their direct or quasi-direct band gap. This is the case for \ch{Cs2GeSnCl6} which has a direct gap of 1.41 eV and and efficiency of 28\%, for \ch{Cs2GeSnBr6}, which has a direct gap of 0.95 eV and an efficiency of 23\%, and for \ch{Rb2CrInI6}, with a direct gap at 0.97 eV and a transition at 1.00 eV. \ch{Cs2GeSnCl6} and \ch{Cs2GeSnBr6} also show optimal charge transport properties due to their low and balanced carrier effective masses (-0.13m$_0$ and 0.20m$_0$ for \ch{Cs2GeSnCl6}, -0.09m$_0$ and 0.14m$_0$ for \ch{Cs2GeSnBr6}, in units of electron masses m$_0$ for holes and electrons, respectively). 

\ch{Rb2CrInI6} has heavier and anisotropic effective masses, but still comparable to that of \ch{Cs2AgBiBr6}. Also some materials with indirect gap show a high efficiency, namely \ch{Cs2AlLiI6}, \ch{Cs2CuInCl6}, \ch{Cs2AgInBr6},\ch{Rb2CrGaI6} and \ch{Cs2CrGaI6}. However these compounds have negative decomposition enthalpy and might not be stable at room temperature. Additionally, their bands are definitely narrower. 

Among the structures with positive decomposition enthalpy , the highest efficiency (17\%) was found for \ch{Cs2CrScI6}, with an indirect gap of 1.37 eV and absorption energy of 1.88 eV. The carriers effective masses are of the order of 0.5 m$_0$, except for holes along the L-$\Gamma$ direction, were m$^*_h$ is -0.87m$_0$.

Finally non symmetry-constrained relaxations show that with a threshold of 0.1 \AA{}  the majority of the structures retains the initial cubic spacegroup \textit{Fm-3m} with the exception of 4 tetragonal structures, namely \ch{Rb2AgIrBr6} (\textit{I4}), \ch{Cs2CuInCl6} (\textit{I4/mmm}), \ch{Rb2CuFeCl6} (\textit{I4/mmm}) and \ch{Rb2CuRuF6} (\textit{I4/mmm}), one monoclinic structure, \ch{Cs2FeRbI6} (\textit{C2/m}), and one triclinic, \ch{Rb2AuMoBr6} (\textit{P-1}).

Due to this second relaxation, the decomposition enthalpy of \ch{Rb2AgInBr6} reaches -0.7 meV/atom and a power conversion efficiency of 17\%.  

\section{Conclusions}
In this work, we screened a large chemical space (7056 compounds) of inorganic halide double perovskites to uncover suitable candidates for photovoltaic applications. We applied a funnel-type approach to identify a pool of potential candidates and then reduce it by successively performing more demanding calculations based on band gap, thermodynamic stability, power conversion efficiency and carrier effective masses. Thereby we employed a state-of-the-art ML approach as a first step to limit the number of expensive band-structure calculation to just the 964 compounds used in training and testing the ML model. This is based on a neural network architecture composed of convolutional and fully connected layers with a periodic table representation of the perovskites. This approach yielded a high accuracy for the prediction of band gaps versus DFT results. The latter were all computed using high accuracy hybrid DFT including spin-orbit coupling in order to ensure high predictivity of our results. We find a number of very high performing compounds---with efficiencies as high as 28\% and very low carrier effective masses (-0.13m$_0$ for holes and 0.20m$_0$ for electrons) for \ch{Cs2GeSnCl6}. Unfortunately, our calculations show that such high performing compounds might only be meta-stable. Among the compounds predicted to be thermodynamically stable, we still find some with efficiencies of up to 17\% (\ch{Rb2AgInBr6}) albeit with worse and more anisotropic effective masses. Notably, when relaxing the strict requirement of cubic symmetry, we find 6 compounds to achieve higher stabilities at lower symmetries.

Thus, while we do find a few novel materials, trade-offs between power conversion efficiency, carrier mobility and (meta-)stability may indeed be unavoidable for this materials class.

\vspace{20px}
\noindent \textbf{Acknowledgements:}
\noindent This work was supported by the Deutsche Forschungsgemeinschaft (DFG, German Research Foundation) under the priority programme SPP 2196, grant No.~RE1509/3701
. HO further thanks the DFG for support within the Heisenberg Program, grant No.~OB425/9-1.

\providecommand*{\mcitethebibliography}{\thebibliography}
\csname @ifundefined\endcsname{endmcitethebibliography}
{\let\endmcitethebibliography\endthebibliography}{}

\end{document}